\documentclass[11pt]{article}

\usepackage{graphicx,amsmath,amsfonts,amssymb,dcolumn,amsthm,slashed,bbm,xspace,pgffor,tikz,mathbbol,latexsym,enumerate,amsmath}
\usepackage{soul}
\usepackage[colorlinks,citecolor=red,urlcolor=cyan,linkcolor=blue]{hyperref}
\usepackage[utf8]{inputenc}
\usepackage[english]{babel}
\usepackage{lmodern}
\usepackage{microtype}
\usepackage{cite}
\usepackage{braket}

\numberwithin{equation}{section}

\begin{document}

\title{\textbf{A relativistic non-perturbative local model of fractons and its non-local perturbative hidden sector}}

\author{\textbf{Rodrigo F.~Sobreiro}\thanks{rodrigo\_sobreiro@id.uff.br}\\\\
\textit{{\small UFF - Universidade Federal Fluminense, Instituto de F\'isica,}}\\
\textit{{\small Av. Litorânea, s/n, 24210-346, Niter\'oi, RJ, Brasil.}}}

\date{}
\maketitle

\begin{abstract}
We construct, from first principles, a covariant local model for scalar fractonic matter coupled to a symmetric tensor gauge field. The free gauge field action is just the one of the Blasi-Maggiore model. The scalar sector, describing fracton charges, is a non-trivial covariant generalization of Pretko's quartic model. Because the model has no quadratic term in the scalar field, a direct perturbative treatment fails. Remarkably, by performing a suitable change of variables, we demonstrate that the action can be driven to a perturbative effective action. However, at the price of carrying non-local interacting terms. We study the perturbative regime of the model first by analyzing the classical field equations and some possible simple solutions, which are in accordance with the expected immobile behavior of fractons. We also derive the fracton dispersion relation and, by playing with the parameters of the model, show that there are at least six distinct phases: one with two massive fractonic modes, one of them being tachyonic; one with massless states associated with a long-range attractive potential; a mixed phase with one massive and one massless state; another one where physical states of the scalar field cannot occur at all in the physical spectrum; a massive phase with states of two different masses; and  second phase where the scalar field cannot be associated with physical particles, in spite of its mass being real. Moreover, we find evidence that fractonic bound states emerge in the model for some of these phases.
\end{abstract}

\maketitle

\newpage

\tableofcontents

\newpage

\section{Introduction}\label{intro}

Fractons are collective excitations that are gaining much attention in key areas of physics  \cite{Vijay:2016phm,Nandkishore:2018sel,Pretko:2020cko}. These fractonic states are supposed to appear in, for example, spin liquids \cite{Nandkishore:2018sel,Pretko:2020cko} and glassy dynamics \cite{Prem:2017qcp,Nandkishore:2018sel,Pretko:2020cko}. These quasi-particles are also related to elasticity theory \cite{Pretko:2017kvd,Nandkishore:2018sel,Pretko:2020cko} and, remarkably, to gravity \cite{Pretko:2017fbf,Nandkishore:2018sel,Pretko:2020cko}. Due to the immobility of fracton charges, fractons also have some relation with Carrollian symmetry \cite{Bergshoeff:2017btm,Guerrieri:2021cdz}, see \cite{Figueroa-OFarrill:2023vbj,Figueroa-OFarrill:2023qty}.

A gauge theoretical approach to describe fractons has been developed in the last few years, see for instance \cite{Pretko:2016lgv,Pretko:2020cko,Bertolini:2022ijb,Blasi:2022mbl,Bertolini:2023juh,Sobreiro:2023fki}. Just like ordinary electrodynamics, fractonic gauge models can be cast in terms of potentials, charges, and gauge principles \cite{Pretko:2016lgv,Pretko:2020cko,Blasi:2022mbl,Bertolini:2022ijb,Bertolini:2023juh, Sobreiro:2023fki}. Particularly, as proved by Blasi and Maggiore \cite{Blasi:2022mbl}, a model of fractons based merely on the gauge principle, special relativity, and polynomial locality supplies a fracton action composed of a covariant fractonic action and the Fierz-Pauli action for spin-2 massless particles \cite{Fierz:1939ix,Gambuti:2021meo}. Thence, the covariant Blasi-Maggiore (BM) model is actually a generalization of the pure fractonic gauge theory, by including gravitational modes on the stage. See also \cite{Bertolini:2023juh,Sobreiro:2023fki} and references therein.

In this work, we construct a gauge invariant model for scalar fractonic matter coupled to a symmetric gauge field. The result is that of the BM model \cite{Blasi:2022mbl,Bertolini:2022ijb,Bertolini:2023juh} and a covariant version of Pretko's scalar model \cite{Pretko:2020cko} for fractons. The scalar action is quartic in the scalar field and does not allow a direct perturbative analysis. However, by performing a suitable change of variables, we attain an effective exact model carrying a quadratic sector. Therefore, a perturbative analysis possibility shows up. The price paid is that the action has non-local interacting terms. 

The field equations are then obtained and discussed. It is first shown that the equation for the fracton accepts a constant field configuration solution if the gauge fractonic field is light-like. Additionally, this static fracton solution induces mass to the gauge field, which behaves as massive waves. Therefore, the charges remain static and immobile even in the presence of fractonic waves. In return, it seems that the waves are slowed down by the presence of the fractonic immobile charges.

We also study the linearized field equations for small fluctuations around the fracton background field. In this case, the fluctuation of the gauge field remains behaving as massive waves if the gauge field is traceless. The fracton fluctuation is ruled by a much more complicated fourth-order free equation but with no influence of the gauge field. Therefore, the background remains fixed and fractonic waves cannot affect them nor the fracton fluctuations. On the other hand, the fractonic waves are slowed down when traveling in the fracton charge background.

From the effective action we are also able to obtain the dispersion relation for fractons. The study of this dispersion relation allows us to investigate the phases of the model by playing with the values of the parameters of the model. We find at least six phases associated with the propagating modes: a massive phase plagued with tachyonic states; a phase with a massive mode and a massless one; a phase with two distinct massive modes; a phase with no fractonic physical modes but with the possibility of bound states; a phase with only one massive mode; and a diluted phase where fractons modes are screened. In all cases, all masses depend on the specific values of the couplings.

In some of these phases, a more rich structure is possibly revealed. For instance, in the diluted phase, the masses of fractons are complex and cannot be associated with physical particles. Nevertheless, gauge invariant operators can be constructed and bound states can be formed \cite{Gribov:1977wm,Sobreiro:2005ec,Dudal:2005na,Dudal:2008sp,Sorella:2010it,Dudal:2011gd,Pereira:2013aza,Capri:2015ixa}. In the bound state favored phase, the propagators are in the form $-p^{-4}$. This propagator is commonly associated with confining potentials \cite{Gribov:1977wm,Sobreiro:2005ec,Dudal:2005na}. 

Interestingly, although we introduce the Higgs potential \cite{Itzykson:1980rh} in the hope of describing an interplay between gap and gapless phases, we find that the masses do not entirely depend on the Higgs parameters. An interesting discussion about the Higgs mechanism in higher rank tensor field theories can be found in \cite{Bulmash:2018lid}.

This work is organized as follows: In Sect.~\ref{FRAC} we define the BM model and construct the covariant gauge invariant fractonic scalar action. In Sect.~\ref{PERT} we obtain the effective model of fractons containing a perturbative piece and a non-local one. In Section \ref{FEQ}, we explore the field equations and discuss some simple solutions. The dispersion relation for fractons is then studied in Section \ref{DR}. Then, finally, in Sect.~\ref{conc} we display our final comments.

\section{Relativistic gauge theory of fractons}\label{FRAC}

We begin by defining the BM model \cite{Blasi:2022mbl,Bertolini:2022ijb} for covariant fractons in Minkowski spacetime with metric given by $\eta_{\mu\nu}\equiv\textrm{diag}(1,-1,-1,-1)$. The fundamental fractonic gauge field in symmetric representation is $h_{\mu\nu}=h_{\nu\mu}$, transforming under gauge transformations as
\begin{equation}
    \delta h_{\mu\nu}=\frac{1}{g}\partial_\mu\partial_\nu\zeta\;.\label{gt1}
\end{equation}
The mass parameter $g$ is introduced in order to enforce $h_{\mu\nu}$ to carry canonical dimension 1 and $\zeta$ to be dimensionless. It will be evident later that $g$ plays the role of a coupling constant between the gauge field and the fractonic charged field. Being symmetric, the gauge field is closely related to the graviton in linearized gravity. It turns out that transformation \eqref{gt1} is just the longitudinal sector of general linear diffeomorphisms in linear gravity. Therefore, the BM model can also be visualized as a special kind of linear gravity where, besides the graviton, fractonic modes are also present. 

In the BM model, the gauge invariant field strength reads
\begin{equation}
    F_{\mu\nu\alpha}=\partial_\mu h_{\nu\alpha}+\partial_\nu h_{\mu\alpha}-2\partial_\alpha h_{\mu\nu}\;,\label{F1}
\end{equation}
which is symmetric with respect to the first two indices, $F_{\mu\nu\alpha}=F_{\nu\mu\alpha}$. The most general quadratic action, invariant under gauge transformations \eqref{gt1}, polynomial in the fields and their derivatives, and depending at most on first order derivatives, is given by
\begin{equation}
    S_h=S_g+S_f\;,\label{acth}
\end{equation}
where $S_g$ stands for the linearized gravity action and $S_f$ is the fracton action, namely
\begin{equation}
    S_g=a\int d^4x\left(\frac{1}{4}F_\mu F^\mu-\frac{1}{6}F_{\mu\nu\alpha}F^{\mu\nu\alpha}\right)\;,\label{actg}
\end{equation}
and
\begin{equation}
    S_f=\frac{b}{6}\int d^4x F_{\mu\nu\alpha}F^{\mu\nu\alpha}\;,\label{actf}
\end{equation}
respectively. The factors $a$ and $b$ are dimensionless parameters and $F_\mu=F^\nu_{\phantom{\nu}\nu\mu}$.

In the present model, dynamical fractonic charges will be described by a dimensionless scalar complex field $\phi$ transforming under gauge transformations by
\begin{eqnarray}
\delta\phi&=&i\zeta\phi\;,\nonumber\\
\delta\phi^*&=&-i\zeta\phi\;.\label{gt2}
\end{eqnarray}

At this point, it is convenient to establish the nomenclature of the fields. In analogy to electrodynamics, the scalar field will be the fracton itself, the \emph{fracton field}, just like the electron is the charge carrier in electrodynamics. The gauge field will be called the \emph{fractonic potential}. Finally, the field strength will be called simply by \emph{fractonic field}, see table \ref{table1}.

\begin{table}[ht]
\centering
\begin{tabular}{|c|c|c|}
	\hline 
Field & Name \\
\hline
$\phi$ & Fracton field \\ 
$h_{\mu\nu}$ & Fractonic potential \\ 
$F_{\mu\nu\nu}$ & Fractonic field\\
\hline 
\end{tabular}
\caption{Nomenclature of the fields adopted in this work.}
\label{table1}
\end{table}

To construct the gauge invariant action of the matter sector we demand, again: gauge invariance, covariance and polynomial locality in the fields and their derivatives. It turns out that the action depends at most on first derivatives will not be possible if we wish to keep gauge invariance. In fact, as discussed in \cite{Pretko:2020cko}, the definition of a covariant derivative is not a trivial task. Inevitably, a non-linear covariant derivative must be defined by
\begin{equation}
D_{\mu\nu}\phi^2=\phi\partial_\mu\partial_\nu\phi-\partial_\mu\phi\partial_\nu\phi-igh_{\mu\nu}\phi^2\;.\label{covd1a}
\end{equation}
whose trace reads
\begin{equation}
    D\phi^2=\phi\Box\phi-\partial_\mu\phi\partial^\mu\phi-igh\phi^2\;,\label{covd1b}
\end{equation} 
with $h=\eta^{\mu\nu}h_{\mu\nu}$ and $\Box=\partial_\mu\partial^\mu$. The covariant derivative indeed is covariant under gauge transformations,
\begin{equation}
    \delta D_{\mu\nu}\phi^2=2i\zeta D_{\mu\nu}\phi^2\;.\label{covd2}
\end{equation}

It turns out that the most simple gauge invariants we can construct out from the fields and $D_{\mu\nu}\phi^2$ are actually quartic in the fields. For instance, we can propose the local action
\begin{eqnarray}
    S_\phi&=&\int d^4x\left\{\frac{e_1}{2}\left[(\phi^*)^2\left(D\phi^2\right)+\phi^2\left(D\phi^2\right)^*\right]+i\frac{e_2}{2}\left[(\phi^*)^2\left(D\phi^2\right)-\phi^2\left(D\phi^2\right)^*\right]+\right.\nonumber\\
    &+&\left.\kappa\left(D^{\mu\nu}\phi^2\right)^*\left(D_{\mu\nu}\phi^2\right)+m^2\phi^*\phi-\frac{\lambda}{2}\left(\phi^*\phi\right)^2\right\}\;,\label{actphi}
\end{eqnarray}
where $e_1$, $e_2$, and $m$ are couplings with mass dimension 2 while $\kappa$ is a dimensionless coupling and $\lambda$ carries mass dimension 4. The Higgs potential is included mostly for generality. Nevertheless, the Higgs potential is an extra feature that could provide a way to interplay between gapless and gap fracton theories, see \cite{Bulmash:2018lid}. In fact, a successful Higgs mechanism could give rise to a mass for the fractonic potential. However, the action \eqref{actphi} has no quadratic term, a property that makes a perturbative analysis very difficult. Moreover, $m^2$ and $\lambda$ are allowed to be negative. In the usual Higgs action, such liberty generates problems like the Hamiltonian being unbounded from below. We expect no such problem here since the Higgs potential does not work alone in the action \eqref{actphi} - The action is composed only of interacting terms. Anyhow, the full gauge invariant action of our model reads
\begin{equation}
S_0=S_h+S_\phi\;.\label{act0}
\end{equation}

For completeness, the dimensions of the fields and parameters are collected in Table \eqref{table2}.
\begin{table}[ht]
\centering
\begin{tabular}{|c|c|c|c|c|c|c|c|c|c|c|c|c|}
	\hline 
Field/Parameter & $\phi$ & $h$ & $F$ & $a$ & $b$ & $g$ & $e_1$ & $e_2$ & $\kappa$ & $m$ & $\lambda$ \\
	\hline 
Dimension & $0$ & $1$ & $2$ & $0$ & $0$ & $1$ & $2$ & $2$ & $0$ & $2$ & $4$\\
\hline 
\end{tabular}
\caption{The canonical dimension of the fields and parameters.}
\label{table2}
\end{table}

It is worth mentioning that action \eqref{actphi} carries terms that generalize some pieces of the non-covariant action proposed in \cite{Pretko:2020cko}. However, a full comparison demands some extra work that is left for future investigation. In here, we focus on working the action \eqref{act0} in order to reveal a perturbative phase and explore the main features of the model.

\section{The hidden perturbative regime}\label{PERT}

In this section we demonstrate that the non-perturbative action \eqref{actphi} has a hidden perturbative regime which will be revealed by a suitable change of variables with a trivial Jacobian. The cost of the transformation is that of the appearance of non-local interactions.

Remarkably, the scalar action of fractons \eqref{actphi} can reveal a hidden perturbative sector with the price of losing locality. To show this property, we first represent the fracton field in polar variables, 
\begin{equation}
\phi=\rho e^{i\theta}\;,\label{polar1}    
\end{equation}
with $\rho$ being a gauge invariant scalar field while the angular field transforms only by an angular translation $\delta\theta=\zeta$. The covariant derivative \eqref{covd1a} becomes now
\begin{equation}
D_{\mu\nu}\phi^2=e^{2i\theta}\left[\rho\partial_\mu\partial_\nu\rho-\partial_\mu\rho\partial_\nu\rho-i\left(gh_{\mu\nu}-\partial_\mu\partial_\nu\theta\right)\rho^2\right]\;.\label{covd3}
\end{equation}
The derivatives of $\theta$ can be absorbed in $h_{\mu\nu}$ by a simple change of variables\footnote{In a similar way that the Nambu-Goldstone massless modes are absorbed by the vector bosons in the Higgs mechanism with a suitable gauge transformation.},
\begin{equation}
    h_{\mu\nu}=\tilde{h}_{\mu\nu}+\frac{1}{g}\partial_\mu\partial_\nu\theta\;,\label{red0}
\end{equation}
which leaves $S_h$ unaltered. Moreover, such a change of variables is just the substitution of the gauge field $h_{\mu\nu}$ by the gauge invariant field $\tilde{h}_{\mu\nu}=h_{\mu\nu}-g^{-1}\partial_\mu\partial_\nu\theta$ appearing in expression \eqref{covd3}. Therefore, the change to polar variables together with transformation \eqref{red0} allow us to write the model in purely gauge invariant field variables. Thence, the expression \eqref{covd3} simplifies to 
\begin{equation}
D_{\mu\nu}\phi^2=e^{2i\theta}D_{\mu\nu}\rho^2\;.\label{covd4}
\end{equation}
Therefore, the scalar action \eqref{actphi} in gauge invariant variables reads
\begin{equation}
    S_\rho=\int d^4x\left[e_1\rho^2\left(\rho\Box\rho-\partial_\mu\rho\partial^\mu\rho\right)+e_2g\tilde{h}\rho^4+\kappa D^*_{\mu\nu}\rho^2 D_{\mu\nu}\rho^2+m^2\rho^2-\frac{\lambda}{2}\rho^4\right]\;.\label{actp}
\end{equation}

Clearly, the polar action \eqref{actp} still has no quadratic terms in the scalar field. However, one can notice that the first term can be cast in the form,
\begin{equation}
    \int d^4x\rho^2\left(\rho\Box\rho-\partial_\mu\rho\partial^\mu\rho\right)=\int d^4x\frac{1}{2}\left(\rho^2\Box\rho^2-\partial_\mu\rho^2\partial^\mu\rho^2\right)\;,
\end{equation} 
suggesting to perform a second change of variables, namely
\begin{equation}
    \Phi=\rho^2\;.\label{polar2}
\end{equation} 
Therefore, a representation with kinematical terms appears,
\begin{equation}
    \rho^2\left(\rho\Box\rho-\partial_\mu\rho\partial^\mu\rho\right)=\frac{1}{2}\left(\Phi\Box\Phi-\partial_\mu\Phi\partial^\mu\Phi\right)\;.
\end{equation}

The problem in the change of variables \eqref{polar2} appears in the third term. Such a change of variables enforces non-locality because
\begin{equation}
    D_{\mu\nu}\rho^2=\frac{1}{2\Phi}\bar{D}_{\mu\nu}\Phi^2=\frac{1}{2}\left(\partial_\mu\partial_\nu\Phi-\frac{1}{\Phi}\partial_\mu\Phi\partial_\nu\Phi-2ig\tilde{h}_{\mu\nu}\Phi\right)\;.
\end{equation}
Notice that the coupling of $\tilde{h}_{\mu\nu}$ and $\Phi$ in the derivative $\bar{D}_{\mu\nu}$ is twice as strong as the coupling with $\rho$ or $\phi$. A natural consequence comes from the fact that $\Phi$ is the product of two $\rho$ fields. In other words, $\Phi$ carries twice the charge of the original fractonic field. Therefore, $\Phi$ can be seen as representing a bound state of two fracton charges. In terms of $\Phi$, action \eqref{actp} reads
\begin{equation}
    S_\Phi=\int d^4x\left(e_1\Phi\Box\Phi+e_2gh\Phi^2+\frac{\kappa}{4\Phi^2} \bar{D}^*_{\mu\nu}\Phi^2 \bar{D}_{\mu\nu}\Phi^2+m^2\Phi-\frac{\lambda}{2}\Phi^2\right)\;.\label{actk1}
\end{equation}
Explicitly, this action is written as
\begin{eqnarray}
    S_\Phi&=&\int d^4x\left[e_1\Phi\Box\Phi+e_2g\tilde{h}\Phi^2+m^2\Phi-\frac{\lambda}{2}\Phi^2+\right.\nonumber\\
    &+&\left.\frac{\kappa}{4} \left(\Phi\Box^2\Phi-\frac{2}{\Phi}\partial_\mu\partial_\nu\Phi\partial^\mu\Phi\partial^\nu\Phi+\frac{1}{\Phi^2}\partial_\mu\Phi\partial^\mu\Phi\partial_\nu\Phi\partial^\nu\Phi+4g^2\tilde{h}_{\mu\nu}\tilde{h}^{\mu\nu}\Phi^2\right)\right]\;.\label{actk2}
\end{eqnarray}

Some redefinitions are helpful at this point. For instance, it is convenient to rescale the scalar field by
\begin{equation}
\Phi\rightarrow \sqrt{2}\kappa^{-1/2}\Phi\;.\label{resc1}
\end{equation} 
Moreover, we redefine the parameters by
\begin{eqnarray}
    e^2&=&\frac{4e_1}{\kappa}\;,\nonumber\\
    \bar{g}&=&\frac{4e_2g}{\kappa}\;,\nonumber\\
    \bar{m}^2&=&\frac{2\sqrt{2}m^2}{\kappa^{1/2}}\;,\nonumber\\
    \mu^4&=&\frac{4\lambda}{\kappa}\;.\label{resc2}
\end{eqnarray}
Thus, by employing the redefinitions \eqref{resc1} and \eqref{resc2} into the action \eqref{actk2}, we finally achieve an effective action for fractons carrying a quadratic part for the fields and non-local interaction terms,
\begin{eqnarray}
    S_\Phi&=&\frac{1}{2}\int d^4x\left\{\Phi\left[\Box(\Box+e^2)-\mu^4\right]\Phi+\left(\bar{g}\tilde{h}+4g^2\tilde{h}_{\mu\nu}\tilde{h}^{\mu\nu}\right)\Phi^2+\frac{2}{\Phi}\partial_\mu\partial_\nu\Phi\partial^\mu\Phi\partial^\nu\Phi-\frac{1}{\Phi^2}\left(\partial_\mu\Phi\partial^\mu\Phi\right)^2+\right.\nonumber\\
    &+&\left.\bar{m}^2\Phi\right\}\;.\label{actk3a}
\end{eqnarray}
This action can be further simplified by a shift in the gauge field (which leaves $S_h$ invariant), namely
\begin{equation}
    \tilde{h}_{\mu\nu}=\bar{h}_{\mu\nu}-\frac{\bar{g}}{8g^2}\eta_{\mu\nu}\;.\label{red1}
\end{equation}
Finally, resulting in
\begin{equation}
    S_\Phi=\frac{1}{2}\int d^4x\left\{\Phi\left[\Box(\Box+e^2)-\bar{\mu}^4\right]\Phi+4g^2\bar{h}_{\mu\nu}\bar{h}^{\mu\nu}\Phi^2+\frac{2}{\Phi}\partial_\mu\partial_\nu\Phi\partial^\mu\Phi\partial^\nu\Phi-\frac{1}{\Phi^2}\left(\partial_\mu\Phi\partial^\mu\right)^2+\bar{m}^2\Phi\right\}\;.\label{actk3}
\end{equation}
with
\begin{equation}
\bar{\mu}^4=\mu^4-\frac{\bar{g}^2}{4g^2}\;.\label{massrenorm0}
\end{equation}

The first term in action \eqref{actk3} is quadratic in the field $\Phi$ defining thus a kinetic sector for fractons. Nevertheless, it has a fourth-order derivative, representing an unusual kinetic term. The second term collects the effective interaction between fractons and the fractonic potential. The third and fourth terms are the non-local interaction terms, bringing highly non-standard features to our model. The last term is just a tadpole term \cite{Itzykson:1980rh}. 

The tadpole in \eqref{actk3} can be eliminated by performing a suitable shift in the scalar field. However, it is evident from the $\sim\bar{h}^2\Phi^2$ coupling that this shift generates a mass term for the gauge field. Thus, one has to choose to deal with a tadpole for the fracton or massive gauge fields. For the moment, we opt for the former.

It is worth mentioning that the action \eqref{actk3} can be directly achieved by the change of variables \eqref{red0} together with the shift \eqref{red1} of the gauge field, namely,
\begin{equation}
h_{\mu\nu}=\bar{h}_{\mu\nu}+\frac{1}{g}\partial_\mu\partial_\nu\theta-\frac{\bar{g}}{8g^2}\eta_{\mu\nu}\;,\label{red1a}
\end{equation}
the redefinitions \eqref{resc2} and the direct change of variables of the scalar field composed of the rescaling \eqref{resc2} and the polar decomposition \eqref{polar1},
\begin{eqnarray}
\phi&=&\sqrt{2}\kappa^{-1}\Phi^{1/2}e^{i\theta}\;,\;,\nonumber\\
\phi^*&=&\sqrt{2}\kappa^{-1}\Phi^{1/2}e^{-i\theta}\;,\label{red2}
\end{eqnarray}
which is simply the composition of \eqref{polar1}, \eqref{polar2}, and \eqref{resc1}. It turns out that the Jacobian associated with transformations \eqref{red1a} and \eqref{red2} is trivial. As a consequence, the present change of variables does not affect the functional measure of the model. This result, although almost trivial, is important and is demonstrated in Appendix \ref{AP1}.

\section{Field equations}\label{FEQ}

The field equations are obtained from the full action \eqref{act0} in terms of the new variables, namely
\begin{equation}
    S_0=S_h+S_\Phi\;.\label{act00}
\end{equation}
We notice that, due to gauge invariance of the action, $S_h(h)=S_h(\tilde{h})$. Moreover, even though $\tilde{h}_{\mu\nu}$ is gauge invariant, a gauge fixing is still needed to provide information about the divergence of the gauge field, according to Helmholtz theorem \cite{Woodside:1999yj}. Employing the general gauge fixing discussed in \cite{Blasi:2022mbl,Bertolini:2023juh}, we can consider the constraint
\begin{equation}
\partial^\mu\tilde{h}_{\mu\nu}+\chi\partial_\nu\tilde{h}=0\;,\label{gf1}
\end{equation}
with $\chi$ being a dimensionless gauge parameter that will be fixed later. The gauge fixing condition \eqref{gf1} must be considered together with the field equations computed directly from the action \eqref{act00}.

The field equation for the scalar field $\Phi$, since it is a real scalar field, is much easier to obtain than for $\phi$. A straightforward computation, originating from the action \eqref{act00}, yields
\begin{equation}
\left[\Box(\Box+e^2)-\bar{\mu}^4\right]\Phi+4g^2\bar{h}_{\mu\nu}\bar{h}^{\mu\nu}\Phi
+\partial_\mu\partial_\nu\left(\frac{1}{\Phi}\partial^\mu\Phi\partial^\nu\Phi\right)
    -\frac{1}{2}\partial_\mu\left(\frac{1}{\Phi^2}\partial^\mu\Phi\partial_\nu\Phi\partial^\nu\Phi\right)=-\frac{1}{2}\bar{m}^2\;.
    \label{feqP}
\end{equation}
For the gauge field, its field equation reads
\begin{equation}
    (b-a)\partial^\alpha F_{\mu\nu\alpha}+a\left[\eta_{\mu\nu}\partial_\alpha F^\alpha-\frac{1}{2}\left(\partial_\mu F_\nu+\partial_\nu F_\mu\right)\right]=-4g^2\bar{h}_{\mu\nu}\Phi^2\;,\label{feqh}
\end{equation}
with $F_{\mu\nu\alpha}$ and $F_\mu$ being computed with $\bar{h}_{\mu\nu}$.

The field equations \eqref{feqP} and \eqref{feqh} consist of non-linear coupled field equations. Therefore, analytic solutions are quite difficult to obtain. Nevertheless, some work can be done in order to, at least, decouple these equations. Moreover, perturbative methods may always be employed.

It turns out that equation \eqref{feqh} can be solved for $\Phi$ after taking its trace in spacetime indices,
\begin{equation}
    \Phi^2=-\frac{(b+2a)}{4g^2\bar{h}}\partial^\mu F_\mu\;.\label{sol1}
\end{equation}
Therefore, knowing $\bar{h}_{\mu\nu}$, the scalar solution is easily found. In terms of $\bar{h}_{\mu\nu}$, solution \eqref{sol1} reads
\begin{equation}
    \Phi^2=\frac{(b+2a)}{2g^2\bar{h}}\left(\Box\bar{h}-\partial^\mu\partial^\nu\bar{h}_{\mu\nu}\right)\;.\label{sol2}
\end{equation}
On the other hand, solution \eqref{sol2} can be put back in equation \eqref{feqh} so we have a decoupled equation for $\bar{h}_{\mu\nu}$,
\begin{equation}
    (b-a)\partial^\alpha F_{\mu\nu\alpha}+a\left[\eta_{\mu\nu}\partial_\alpha F^\alpha-\frac{1}{2}\left(\partial_\mu F_\nu+\partial_\nu F_\mu\right)\right]=2(b+2a)\frac{\left(\partial^\mu\partial^\nu\bar{h}_{\mu\nu}-\Box\bar{h}\right)}{\bar{h}}\bar{h}_{\mu\nu}\;,\label{feqh2}
\end{equation}
Still a highly non-linear equation, but a single equation for $\bar{h}_{\mu\nu}$ and independent of the coupling $g$. In principle, its solution can be inserted in the $\Phi$ equation \eqref{feqP} and solved for the scalar field. Moreover, since equation \eqref{feqh2} has a linear part, it can be solved with the help of perturbative techniques. 

The final step in the gauge field equation is to use the gauge fixing equation \eqref{gf1} in \eqref{feqh2}.
It becomes,
\begin{equation}
    (b-a)\Box\bar{h}_{\mu\nu}+\left\{\left[\left(b-2a\right)\chi-a\right]\partial_\mu\partial_\nu+a\left(\chi+1\right)\eta_{\mu\nu}\Box\right\}\bar{h}=-(b+2a)\left(\chi+1\right)\frac{\Box\bar{h}}{\bar{h}}\bar{h}_{\mu\nu}\;,\label{feqh3b}
\end{equation}

We remark that, according to \cite{Blasi:2022mbl,Bertolini:2023juh}, the situations $\chi+1=0$, $b-a=0$, and $b+2a=0$ are forbidden. Any other options are at our disposal for simplifications. Let us consider then the gauge
\begin{equation}
    \chi=\frac{a}{b-2a}\;.\,\label{gf2}
\end{equation}
so the mixed derivative term is killed. Thus, equation \eqref{feqh3b} reduces to
\begin{equation}
    \Box\left[\bar{h}_{\mu\nu}+\frac{a}{(b-2a)}\bar{h}\eta_{\mu\nu}\right]=-\frac{(b+2a)(b-a)}{(b-2a)}\frac{\Box\bar{h}}{\bar{h}}\bar{h}_{\mu\nu}\;,\label{feqh3c}
\end{equation}

We remark that the scalar field in terms of equation \eqref{sol2}, in the gauge \eqref{gf2}, now reads
\begin{equation}
    \Phi^2=\frac{(b+2a)(b-a)}{(b-2a)}\frac{1}{2g^2\bar{h}}\Box\bar{h}\;.\label{sol3}
\end{equation}

The set of equations \eqref{feqP} and \eqref{feqh3c} are the final equations we are interested in.

\subsection{Massive waves and constant charged field}

Let us consider the simplest solution, a constant scalar field. Equation \eqref{feqP} allows such a situation if, 
\begin{equation}
    \Phi_o=\frac{\bar{m}^2}{2\bar{\mu}^4}\;.\label{C1}
\end{equation}
while the gauge field is light-like,
\begin{equation}
\bar{h}_{\mu\nu}\bar{h}^{\mu\nu}=0\;.\label{C2}
\end{equation}
Combining the solution \eqref{C1} and  equations \eqref{feqh3c} and \eqref{sol3}, one easily finds that the trace $\bar{h}$ must obey a massive wave equation of the form
\begin{equation}
    \left(\Box+m^2_h\right)\bar{h}=0\;,\label{wave1}
\end{equation}
with the mass of the gauge field being given by
\begin{equation}
    m_h^2=\frac{g^2\bar{m}^4}{2\bar{\mu}^8}\;.\label{mh1}
\end{equation}

The construction of a light-like massive wave is not difficult. For instance, for plane waves, we have the simple solution
\begin{equation}
    \bar{h}_{\mu\nu}=H_{\mu\nu}\left[A\cos\left(p_\mu x^\mu\right)+B\sin\left(p_\mu x^\mu\right)\right]\;,\label{C3}
\end{equation}
with the condition that the amplitude tensor $H_{\mu\nu}$ is a light-like constant field while $A$ and $B$ are integration constants. Moreover, in order for the expression to actually satisfy equation \eqref{wave1}, the gauge field must obey the usual massive dispersion relation, 
\begin{equation}
    p_\mu p^\mu-m_h^2=0\;,
\end{equation}
with the condition that.

This is a curious solution. It seems that a constant fractonic configuration is persistent, in the sense that not even fractonic waves can induce dynamics to the charges, remaining immobile. Moreover, this configuration induces mass to the fracton waves. Apparently, the tendency of immobility of fractons enforces fractonic waves to travel slower than expected. 

Another way to interpret this is that there is a critical situation for a specific value of the scalar field that induces mass to the gauge field. This is actually evident from the action \eqref{feqP}: to eliminate the tadpole, we need to establish that there is a background value for the scalar field which is exactly the solution \eqref{C1}. The performance of this shift in the action \eqref{feqP} not only eliminates the tadpole, but also induces a mass term for the gauge field. It seems thus that the mass gap for the gauge field is also a persistent effect.

\subsection{Linearized equations}

let us linearize the field equations by considering small perturbations around the background fracton field\footnote{We remark that, if the solution \eqref{C1} is assumed for the fracton field, a constant field configuration for the gauge field is ruled out due to the mass term in Equation \eqref{wave1}. Therefore, a possible background for the gauge field must be viewed as an external field.},
\begin{eqnarray}
\bar{h}_{\mu\nu}&=&\mathrm{h}_{\mu\nu}\;,\nonumber\\
\Phi&=&\Phi_o+\varphi\label{diag1}
\end{eqnarray}
with $c_{\mu\nu}$ and $\mathrm{h}_{\mu\nu}$ and $\varphi$ being small fluctuations. The corresponding linearized field equations obtained from equations \eqref{feqP} and \eqref{feqh}, in the gauge \eqref{gf2}, are
\begin{eqnarray}
    \left[\Box(\Box+e^2)-\bar{\mu}^4\right]\varphi&=&0\;,\nonumber\\
    \Box\left[\mathrm{h}_{\mu\nu}+\frac{a}{(b-2a)}\eta_{\mu\nu}\mathrm{h}\right]+m_h^2\mathrm{h}_{\mu\nu}&=&0\;,\label{feqrad1}
\end{eqnarray}
Therefore, the fracton fluctuation is actually dynamical but free, obeying a fourth-order massive linear equation. The gauge field obeys a kind of massive wave equation. In fact, by requiring that the gauge field is traceless, we have that equations \eqref{feqrad1} become
\begin{eqnarray}
    \left[\Box(\Box+e^2)-\bar{\mu}^4\right]\varphi&=&0\;,\nonumber\\
\left(\Box+m_h^2\right)\mathrm{h}_{\mu\nu}&=&0\;.\label{feqrad2}
\end{eqnarray}
The equations are thus completely decoupled. 

We can thus infer that the fracton background still persists to the existence of the fractonic waves. Moreover, fracton fluctuations are allowed as dynamical configurations as long as they are free, \emph{i.e.}, they do not care about the waves. Nevertheless, the waves are massive and the mass is induced by the fracton background. Thence, even though fractons are not influenced by the fractonic waves, the fractonic waves are slowed down in the fracton charged sea.

\section{The fracton dispersion relation}\label{DR}

The free fracton equation, obtained by turning off all non-linear pieces of equation \eqref{feqP}, reads
\begin{equation}
\Box\left[(\Box+e^2)-\bar{\mu}^4\right]\Phi=0\;.\label{drel1}
\end{equation}
where the tadpole term was eliminated by the shift
\begin{equation}
\Phi\longrightarrow\Phi+\frac{\bar{m}^2}{2\bar{\mu}^4}\;.
\end{equation}
Transforming equation \eqref{drel1} in momentum space, we directly obtain the dispersion relation of fractons,
\begin{equation}
p^2(p^2-e^2)-\bar{\mu}^4=0\;,\label{drel2}
\end{equation}
which can be decomposed in terms of two different massive states,
\begin{equation}
    \left(p^2-e^2_+\right)\left(p^2-e^2_-\right)=0\;\label{drel3}
\end{equation}
with masses given by
\begin{equation}
e_\pm^2=\frac{e^2\pm\sqrt{e^4+4\bar{\mu}^4}}{2}\;.\label{poles1}
\end{equation}
The analysis of the fracton spectrum clearly depends on the parameters $e$ and $\bar{\mu}$, defined in \eqref{resc2} and \eqref{massrenorm0}. 

Let us start the scrutinization of the fracton spectrum generated by the dispersion relation \eqref{drel2} by looking at the high energy regime and considering real masses $e_\pm$. The dispersion relation \eqref{drel2} becomes $p^4=0$. Or, from \eqref{drel3}, two massless states. From the quantum field theoretical point of view, this situation corresponds to a propagator of the form $\braket{\Phi|\Phi}\sim p^{-4}$. Although such a propagator is usually unrelated to physical states, it is associated with a long-range potential. Thus, at very high energies, bound states may appear due to this potential.

We split the spectrum analysis into three parts: the case where $\bar{\mu}^4$ is positive definite; the case $\bar{\mu}^4=0$; and $\bar{\mu}^4<0$. In all cases, we keep $e^2$ positive definite. Thence, the analysis is subjected to vary along the possible values of
\begin{equation}
    \bar{\mu}^2=\frac{4}{\kappa}\left(\lambda-\frac{e^2_2}{\kappa}\right)\;.\label{mubar1}
\end{equation}
It turns out that we have six different possibilities. The first three are:
\begin{itemize}
    \item \emph{Massive phase plagued with tachyons}: First we consider $\bar{\mu}^4>0$. In this regime, we have the situation where fractons occur generally with two distinct masses given by the \eqref{poles1}. However, only one of these masses is positive,
\begin{eqnarray}
    e^2_+&>&0\;,\nonumber\\
    e^2_-&<&0\;.\label{tach1}
\end{eqnarray}
Therefore, the states with mass $e_-^2$ are directly identified with tachyons. It seems thus, at first sight, that the case of positive $\bar{\mu}^4$ is ruled out for the present model.

\item \emph{Tachyon-free massive phase $\#1$}: Considering now $\bar{\mu}^4=0$, which means that, as seen from expression \eqref{mubar1}, some parameters of the model must obey a constraint, namely
\begin{equation}
e^2_2=\kappa\lambda\;.\label{cond3}
\end{equation}
From where it is clear that if we also choose $\lambda=0$, thence $e_2$ must necessarily vanish. 

Making $\bar{\mu}^4$ vanish in expression \eqref{poles1} results in
\begin{eqnarray}
    e^2_+&=&e^2\;,\nonumber\\
    e^2_-&=&0\;.\label{notach1}
\end{eqnarray}
In this phase, we have massive and massless modes. Moreover, no tachyons show up. We notice that this phase is independent of the Higgs potential, meaning that the fracton mass gap can be generated from an entirely different mechanism.

\item \emph{Bound states favoring phase}: In the special case that $e^2$ also vanishes, we find a situation similar to the high energy limit, where only massless states are allowed and the propagator behaves as $\sim p^{-4}$. Thus, bound states may be favored in the massless case, without the necessity of increasing the energy.

\end{itemize}

The cases with $\bar{\mu}^4<0$ are also interesting to take a look at because they generate at least three more possible phases. Before analyzing these phases, we first observe that to $\bar{\mu}^4<0$ be true, it is enough, but mandatory, to set $\lambda=0$ (see expression \eqref{mubar1}). Thence, the Higgs potential, again, has no effect in these phases. 

In order to make the results clearer, let us rewrite the masses \eqref{poles1} with the explicit minus sign,
 \begin{equation}
    e^2_\pm=\frac{e^2\pm\sqrt{e^4-4M^2}}{2}\;,\label{poles2}
\end{equation}
with $M^4=|\bar{\mu}^2|$.

The three extra phases are:
\begin{itemize}
    \item \emph{Tachyon-free massive phase $\#2$}: This situation occurs for the case $e^4>4M^4$, generating two distinct real masses
    \begin{eqnarray}
        e^2_+=\frac{e^2+\sqrt{e^4-4M^2}}{2}\;,\nonumber\\
        e^2_-=\frac{e^2-\sqrt{e^4-4M^2}}{2}\;,\label{poles3}
    \end{eqnarray}
    
    Therefore, we have two fractonic states with different non-vanishing masses. Again, no tachyons show up.

    \item \emph{Diluted phase}: The case $e^4<4M^4$ (including $e^2=0$) generates a phase where we have two distinct complex masses. However, although the propagator cannot be associated with a physical particle, one is still free to find nontrivial gauge invariant operators that may propagate with real massive poles, suggesting a richer structure hidden in the model. See for instance \cite{Gribov:1977wm,Sobreiro:2005ec,Dudal:2005na,Dudal:2008sp,Sorella:2010it,Dudal:2011gd,Capri:2012cr,Pereira:2013aza,Capri:2015ixa} in the context of quark-gluon confinement. The complex poles indicate that pure fractonic modes cannot occur as physical observables at this regime. In other words, they somehow dilute in the media and disappear, favoring more complex structures.

     \item \emph{Diluted phase with a single massive phase}: It is worth mentioning the special case where $e^4=4\mu^4$, which provides two states with equal masses characterized by the single pole of second-order $e_+^2=e_-^2=e^2/2$. In this case, the decomposition \eqref{drel2} of the propagator is not valid and we have actually one single massive state. Nevertheless, due to the pole be of second order, this state may not correspond to an asymptotic state, meaning that this phase can also be viewed as a special case of the diluted phase.

\end{itemize}

The most promising situation is thus $\bar{\mu}^4=0$, where all states are well-defined. The case $\bar{\mu}^4<0$ is also interesting due to the presence of distinct fractonic phases. The worse situation is then $\bar{\mu}^4>0$, where the possibility of tachyons arises.

\section{Conclusions}\label{conc}

We have constructed a covariant gauge theory for scalar dynamical fractons interacting via a symmetric tensor gauge field mediator. The model, constructed based on purely gauge principles, is composed of the Blasi-Maggiore action for the tensor gauge field \cite{Blasi:2022mbl,Bertolini:2022ijb,Bertolini:2023juh} and a covariant generalization of Pretko's action \cite{Pretko:2020cko} for scalar fractonic matter. The model we constructed, just like the non-covariant proposal of \cite{Pretko:2020cko}, is a quartic model on the scalar fields. As a consequence, no direct perturbative analysis is at our disposal. Nevertheless, we have found a way to obtain an effective action carrying a quadratic piece in the fields, therefore revealing a perturbative sector for the model. 

We have discussed the classical field equations of the model. We found an exact solution of constant fracton field coexisting with fractonic light-like massive waves. This solution is consistent with the natural immobility tendency of fractons. It seems thus that the fracton constant configuration is persistent, and its effect on fractonic waves is that of generating mass for these waves. 

Going to the linearized equations around the fracton background, we found that the gauge field fluctuation remains behaving as massive waves while the background remains solid. The fracton fluctuations also do not interact with the waves, obeying a free equation of fourth order.

By studying the dispersion relation of the effective fractonic matter field, even in a zero temperature environment, we were able to find at least six fractonic phases based on the values of some parameters of the model. Our results show that the effective model obtained can travel between phases where no fractons are allowed, massless fractons appear, a mixed phase of massless fractons and massive ones co-exist, a full massive fractonic phase emerges, and, in one specific case, tachyonic modes show up. In these phases, there are also evidences that fractonic bound states may appear, a very welcome feature for a fractonic model. Our conclusion is that the model is suitable to describe not only gapless fractonic models and massive fractons, but also mixed phases, and a diluted phase where no fracton modes appear at all.

In summary, we have a covariant theory more connected with Pretko's model \cite{Pretko:2020cko}. A quartic model with no quadratic terms for the fractons. On the other hand, we have an effective model with quadratic terms but carrying non-local interacting terms for the fracton field. In both situations, non-trivial analysis is required. In fact, still concerning Pretko's model, another point to be investigated is the non-relativistic limit of the model constructed in the present letter. For instance, applying the standard methods of \cite{Bergshoeff:2017btm,Hansen:2020pqs,Guerrieri:2020vhp}, one could relate the covariant actions introduced here with the non-covariant fracton theory proposed in \cite{Pretko:2020cko}. It would be also interesting to perform not only standard loop expansion computations \cite{Itzykson:1980rh} but also do it within finite temperature formalism \cite{Kapusta:1989tk}. Thence, a study of phase transitions between the phases we inferred could be explored. It is worth mentioning that the same procedure here developed can be performed with first order variables, just like it was done in \cite{Sobreiro:2023fki}. In the refereed work, a fracton phase where ordinary electrodynamics is indistinguishable from fracton dynamics was found. Therefore, an analysis in the representation discussed in \cite{Sobreiro:2023fki} can be of physical relevance. Another interesting possibility is to explore the model as a toy model for quark-gluon confinement. Such an idea arises from the fact that the propagators obtained in our effective theory are quite similar to those studied in improved quantized Yang-Mills theories by taking into account Gribov copies and non-perturbative effects of QCD \cite{Gribov:1977wm,Sobreiro:2005ec,Dudal:2005na,Dudal:2008sp,Dudal:2011gd,Pereira:2013aza,Capri:2015ixa}. As a final comment, curiously, the word ``fracton'' was first used in a quark-gluon confinement theoretical scenario years ago in Khoplov's ideas connecting confinement to fractional charged particles \cite{Khlopov:1981wm,Khlopov:1999rs,Khlopov:2012lua}.

\section*{Acknowledgements}

I am grateful to L.~E.~Oxman for many useful and clarifying discussions. This study was financed in part by The Coordena\c c\~ao de Aperfei\c coamento de Pessoal de N\'ivel Superior - Brasil (CAPES) - Finance Code 001. 

\appendix

\section{The trivial Jacobian}\label{AP1}

The change of variables (see expressions \eqref{red0} and \eqref{red2}), up to irrelevant factors, is (see expressions \eqref{red0} and \eqref{red2})
\begin{eqnarray}
\phi&=&\Phi^{1/2}e^{i\theta}\;,\;,\nonumber\\
\phi^*&=&\Phi^{1/2}e^{-i\theta}\;,\nonumber\\
h_{\mu\nu}&=&\bar{h}_{\mu\nu}+\frac{1}{g}\partial_\mu\partial_\nu\theta-\frac{\bar{g}}{8g^2}\eta_{\mu\nu}\;,
\label{red3}
\end{eqnarray}
from which we compute the elements of the transformation matrix,
\begin{eqnarray}
    \frac{\partial\phi}{\partial\Phi}&=&\Phi^{-1/2}e^{i\theta}\;,\nonumber\\
    \frac{\partial\phi}{\partial\theta}&=&i\Phi^{1/2}e^{i\theta}\;,\nonumber\\
    \frac{\partial\phi}{\partial h_{\alpha\beta}}&=&0\;,\nonumber\\
    \frac{\partial\phi^*}{\partial\Phi}&=&\Phi^{-1/2}e^{-i\theta}\;,\nonumber\\
    \frac{\partial\phi^*}{\partial\theta}&=&-i\Phi^{1/2}e^{-i\theta}\;,\nonumber\\
    \frac{\partial\phi^*}{\partial h_{\alpha\beta}}&=&0\;,\nonumber\\
     \frac{\partial h_{\mu\nu}}{\partial\Phi}&=&0\;,\nonumber\\
     \frac{\partial h_{\mu\nu}}{\partial\theta}&=&\frac{1}{g}\partial_\mu\partial_\nu\;,\nonumber\\
     \frac{\partial h_{\mu\nu}}{\partial\bar{h}_{\alpha\beta}}&=&\frac{1}{2}\left(\delta_\mu^\alpha\delta_\nu^\beta+\delta_\mu^\beta\delta_\nu^\alpha\right)\;.
\end{eqnarray}
Thus, the corresponding Jacobian $J$ is easily computed,
\begin{equation}
    J=\begin{vmatrix}
\Phi^{-1/2}e^{i\theta} & i\Phi^{1/2}e^{i\theta} & 0\\
\Phi^{-1/2}e^{-i\theta} & -i\Phi^{1/2}e^{-i\theta} & 0\\
0 & \frac{1}{g}\partial_\mu\partial_\nu & \frac{1}{2}\left(\delta_\mu^\alpha\delta_\nu^\beta+\delta_\mu^\beta\delta_\nu^\alpha\right)
\end{vmatrix}=-i\Big|\left(\delta_\mu^\alpha\delta_\nu^\beta+\delta_\mu^\beta\delta_\nu^\alpha\right)\Big|=-20i\;.
\end{equation}
This result establishes that the transformation does not affect the functional generators of the model.

\bibliography{BIB}

\providecommand{\href}[2]{#2}\begingroup\raggedright\begin{thebibliography}{10}

\bibitem{Vijay:2016phm}
S.~Vijay, J.~Haah, and L.~Fu, ``{Fracton Topological Order, Generalized Lattice Gauge Theory and Duality}''. \href{http://dx.doi.org/10.1103/PhysRevB.94.235157}{{\em Phys. Rev. B} {\bfseries 94} no.~23, (2016) 235157}.

\bibitem{Nandkishore:2018sel}
R.~M. Nandkishore and M.~Hermele, ``{Fractons}''. \href{http://dx.doi.org/10.1146/annurev-conmatphys-031218-013604}{{\em Ann. Rev. Condensed Matter Phys.} {\bfseries 10} (2019) 295--313}.

\bibitem{Pretko:2020cko}
M.~Pretko, X.~Chen, and Y.~You, ``{Fracton Phases of Matter}''. \href{http://dx.doi.org/10.1142/S0217751X20300033}{{\em Int. J. Mod. Phys. A} {\bfseries 35} no.~06, (2020) 2030003}.

\bibitem{Prem:2017qcp}
A.~Prem, J.~Haah, and R.~Nandkishore, ``{Glassy quantum dynamics in translation invariant fracton models}''. \href{http://dx.doi.org/10.1103/PhysRevB.95.155133}{{\em Phys. Rev. B} {\bfseries 95} no.~15, (2017) 155133}.

\bibitem{Pretko:2017kvd}
M.~Pretko and L.~Radzihovsky, ``{Fracton-Elasticity Duality}''. \href{http://dx.doi.org/10.1103/PhysRevLett.120.195301}{{\em Phys. Rev. Lett.} {\bfseries 120} no.~19, (2018) 195301}.

\bibitem{Pretko:2017fbf}
M.~Pretko, ``{Emergent gravity of fractons: Mach\textquoteright{}s principle revisited}''. \href{http://dx.doi.org/10.1103/PhysRevD.96.024051}{{\em Phys. Rev. D} {\bfseries 96} no.~2, (2017) 024051}.

\bibitem{Bergshoeff:2017btm}
E.~Bergshoeff, J.~Gomis, B.~Rollier, J.~Rosseel, and T.~ter Veldhuis, ``{Carroll versus Galilei Gravity}''. \href{http://dx.doi.org/10.1007/JHEP03(2017)165}{{\em JHEP} {\bfseries 03} (2017) 165}.

\bibitem{Guerrieri:2021cdz}
A.~Guerrieri and R.~F. Sobreiro, ``{Carroll limit of four-dimensional gravity theories in the first order formalism}''. \href{http://dx.doi.org/10.1088/1361-6382/ac345f}{{\em Class. Quant. Grav.} {\bfseries 38} no.~24, (2021) 245003}.

\bibitem{Figueroa-OFarrill:2023vbj}
J.~Figueroa-O'Farrill, A.~P\'erez, and S.~Prohazka, ``{Carroll/fracton particles and their correspondence}''. \href{http://dx.doi.org/10.1007/JHEP06(2023)207}{{\em JHEP} {\bfseries 06} (2023) 207}.

\bibitem{Figueroa-OFarrill:2023qty}
J.~Figueroa-O'Farrill, A.~P\'erez, and S.~Prohazka, ``{Quantum Carroll/fracton particles}''. \href{http://dx.doi.org/10.1007/JHEP10(2023)041}{{\em JHEP} {\bfseries 10} (2023) 041}.

\bibitem{Pretko:2016lgv}
M.~Pretko, ``{Generalized Electromagnetism of Subdimensional Particles: A Spin Liquid Story}''. \href{http://dx.doi.org/10.1103/PhysRevB.96.035119}{{\em Phys. Rev. B} {\bfseries 96} no.~3, (2017) 035119}.

\bibitem{Bertolini:2022ijb}
E.~Bertolini and N.~Maggiore, ``{Maxwell theory of fractons}''. \href{http://dx.doi.org/10.1103/PhysRevD.106.125008}{{\em Phys. Rev. D} {\bfseries 106} no.~12, (2022) 125008}.

\bibitem{Blasi:2022mbl}
A.~Blasi and N.~Maggiore, ``{The theory of symmetric tensor field: From fractons to gravitons and back}''. \href{http://dx.doi.org/10.1016/j.physletb.2022.137304}{{\em Phys. Lett. B} {\bfseries 833} (2022) 137304}.

\bibitem{Bertolini:2023juh}
E.~Bertolini, A.~Blasi, A.~Damonte, and N.~Maggiore, ``{Gauging fractons and linearized gravity}''. \href{http://dx.doi.org/10.3390/sym15040945}{{\em Symmetry} {\bfseries 15} (2023) 945}.

\bibitem{Sobreiro:2023fki}
R.~F. Sobreiro, ``{Unifying fractons, gravitons and photons from a gauge theoretical approach}''. \url{https://arxiv.org/abs/2306.04589}.

\bibitem{Fierz:1939ix}
M.~Fierz and W.~Pauli, ``{On relativistic wave equations for particles of arbitrary spin in an electromagnetic field}''. \href{http://dx.doi.org/10.1098/rspa.1939.0140}{{\em Proc. Roy. Soc. Lond. A} {\bfseries 173} (1939) 211--232}.

\bibitem{Gambuti:2021meo}
G.~Gambuti and N.~Maggiore, ``{Fierz\textendash{}Pauli theory reloaded: from a theory of a symmetric tensor field to linearized massive gravity}''. \href{http://dx.doi.org/10.1140/epjc/s10052-021-08962-8}{{\em Eur. Phys. J. C} {\bfseries 81} no.~2, (2021) 171}.

\bibitem{Gribov:1977wm}
V.~N. Gribov, ``{Quantization of Nonabelian Gauge Theories}''. \href{http://dx.doi.org/10.1016/0550-3213(78)90175-X}{{\em Nucl. Phys.} {\bfseries B139} (1978) 1}.
[,1(1977)].

\bibitem{Sobreiro:2005ec}
R.~F. Sobreiro and S.~P. Sorella, ``{Introduction to the Gribov ambiguities in Euclidean Yang-Mills theories}''. in {\em {13th Jorge Andre Swieca Summer School on Particle and Fields Campos do Jordao, Brazil, January 9-22, 2005}}.
\newblock
2005.
\newblock

\bibitem{Dudal:2005na}
D.~Dudal, R.~F. Sobreiro, S.~P. Sorella, and H.~Verschelde, ``{The Gribov parameter and the dimension two gluon condensate in Euclidean Yang-Mills theories in the Landau gauge}''.
\href{http://dx.doi.org/10.1103/PhysRevD.72.014016}{{\em Phys. Rev.} {\bfseries D72} (2005) 014016}.

\bibitem{Dudal:2008sp}
D.~Dudal, J.~A. Gracey, S.~P. Sorella, N.~Vandersickel, and H.~Verschelde, ``{A Refinement of the Gribov-Zwanziger approach in the Landau gauge: Infrared propagators in harmony with the lattice results}''. \href{http://dx.doi.org/10.1103/PhysRevD.78.065047}{{\em Phys. Rev. D} {\bfseries 78} (2008) 065047}.

\bibitem{Sorella:2010it}
S.~P. Sorella, ``{Gluon confinement, i-particles and BRST soft breaking}''. \href{http://dx.doi.org/10.1088/1751-8113/44/13/135403}{{\em J. Phys. A} {\bfseries 44} (2011) 135403}.

\bibitem{Dudal:2011gd}
D.~Dudal, S.~P. Sorella, and N.~Vandersickel, ``{The dynamical origin of the refinement of the Gribov-Zwanziger theory}''.
\href{http://dx.doi.org/10.1103/PhysRevD.84.065039}{{\em Phys. Rev.} {\bfseries D84} (2011) 065039}.

\bibitem{Pereira:2013aza}
A.~D. Pereira and R.~F. Sobreiro, ``{On the elimination of infinitesimal Gribov ambiguities in non-Abelian gauge theories}''.
\href{http://dx.doi.org/10.1140/epjc/s10052-013-2584-6}{{\em Eur. Phys. J.} {\bfseries C73} (2013) 2584}.

\bibitem{Capri:2015ixa}
M.~A.~L. Capri, D.~Dudal, D.~Fiorentini, M.~S. Guimaraes, I.~F. Justo, A.~D. Pereira, B.~W. Mintz, L.~F. Palhares, R.~F. Sobreiro, and S.~P. Sorella, ``{Exact nilpotent nonperturbative BRST symmetry for the Gribov-Zwanziger action in the linear covariant gauge}''.
\href{http://dx.doi.org/10.1103/PhysRevD.92.045039}{{\em Phys. Rev.} {\bfseries D92} no.~4, (2015) 045039}.

\bibitem{Itzykson:1980rh}
C.~Itzykson and J.~B. Zuber, {\em {Quantum Field Theory}}.
\newblock International Series In Pure and Applied Physics. McGraw-Hill, New York, 1980.
\newblock
\url{http://dx.doi.org/10.1063/1.2916419}.
\newblock

\bibitem{Bulmash:2018lid}
D.~Bulmash and M.~Barkeshli, ``{The Higgs Mechanism in Higher-Rank Symmetric $U(1)$ Gauge Theories}''. \href{http://dx.doi.org/10.1103/PhysRevB.97.235112}{{\em Phys. Rev. B} {\bfseries 97} no.~23, (2018) 235112}.

\bibitem{Woodside:1999yj}
D.~A. Woodside, ``{Uniqueness theorems for classical four-vector fields in Euclidean and Minkowski spaces}''. \href{http://dx.doi.org/10.1063/1.533007}{{\em J. Math. Phys.} {\bfseries 40} (1999) 4911--4943}.

\bibitem{Capri:2012cr}
M.~A.~L. Capri, D.~Dudal, A.~J. Gomez, M.~S. Guimaraes, I.~F. Justo, and S.~P. Sorella, ``{A study of the Higgs and confining phases in Euclidean SU(2) Yang-Mills theories in 3d by taking into account the Gribov horizon}''. \href{http://dx.doi.org/10.1140/epjc/s10052-013-2346-5}{{\em Eur. Phys. J. C} {\bfseries 73} no.~3, (2013) 2346}.

\bibitem{Hansen:2020pqs}
D.~Hansen, J.~Hartong, and N.~A. Obers, ``{Non-Relativistic Gravity and its Coupling to Matter}''. \href{http://dx.doi.org/10.1007/JHEP06(2020)145}{{\em JHEP} {\bfseries 06} (2020) 145}.

\bibitem{Guerrieri:2020vhp}
A.~Guerrieri and R.~F. Sobreiro, ``{Non-relativistic limit of gravity theories in the first order formalism}''. \href{http://dx.doi.org/10.1007/JHEP03(2021)104}{{\em JHEP} {\bfseries 03} (2021) 104}.

\bibitem{Kapusta:1989tk}
J.~I. Kapusta, {\em {Finite Temperature Field Theory}}.
\newblock Cambridge Monographs on Mathematical Physics. Cambridge University Press, Cambridge, 1989.

\bibitem{Khlopov:1981wm}
M.~Y. Khlopov, ``{FRACTIONALLY CHARGED PARTICLES AND CONFINEMENT OF QUARKS}''. {\em Pisma Zh. Eksp. Teor. Fiz.} {\bfseries 33} (1981) 170--173.

\bibitem{Khlopov:1999rs}
M.~Y. Khlopov, \href{http://dx.doi.org/10.1142/11875}{{\em {Cosmoparticle physics}}}.
\newblock World Scientific, Singapore, 1999.

\bibitem{Khlopov:2012lua}
M.~Khlopov, \href{http://dx.doi.org/10.1007/978-1-907343-72-8}{{\em {Fundamentals of Cosmic Particle Physics}}}.
\newblock Cambridge International Science Publishing, Cambridge, UK, 2012.

\end{thebibliography}\endgroup
\bibliographystyle{utphys2}

\end{document}